\documentclass[preprint,aps,showpacs]{revtex4}
\usepackage{graphicx}
\usepackage{latexsym}
\begin{document}
%
\title{Exact scalings in competitive growth models}
\author{L. A. Braunstein$^{1,3}$  and Chi-Hang Lam$^2$}

\affiliation{$^1$Departamento de F\'{\i}sica, Facultad de Ciencias Exactas y
Naturales\\ Universidad Nacional de Mar del Plata\\ Funes 3350,
$7600$ Mar del Plata, Argentina\\
$^3$Center for Polymer Studies and Department of Physics\\
Boston University, Boston, MA 02215, USA \\
$^2$ Department of Applied Physics, Hong Kong Polytechnic University, Hung
Hom, Hong Kong}


\begin{abstract}
A competitive growth model (CGM) describes aggregation of a single
type of particle under two distinct growth rules with occurrence
probabilities $p$ and $1-p$. We explain the origin of scaling
behaviors of the resulting surface roughness with respect to $p$ for
two CGMs which describe random deposition (RD) competing with ballistic
deposition (BD) and RD competing with the Edward Wilkinson
(EW) growth rule. Exact scaling exponents are derived and are in
agreement with previously conjectured values
\cite{albano,albanojpa}. Using this analytical result we are able to
derive theoretically the scaling behaviors of the coefficients of the
continuous equations that describe their universality classes.  We also
suggest that, in some CGM, the $p-$dependence on the coefficients of
the continuous equation that represents the universality class can be
non trivial. In some cases the process cannot be represented by a
unique universality class. In order to show this we introduce a CGM
describing RD competing with a constrained EW (CEW) model. This CGM
show a transition in the scaling exponents from RD to a
Kardar-Parisi-Zhang behavior when $p \to 0$ and to a Edward Wilkinson
one when $p \to 1$. Our simulation results are in excellent agreement with
the analytic predictions.
\end{abstract}

\maketitle
\section{Introduction}
Evolving interfaces or surfaces are of great interest due to their
potential technological applications. These interfaces can be found in
many physical, chemical and biological processes. Examples include film
growth either by vapour deposition or chemical deposition
\cite{family}, bacterial colony growth \cite{albanobact} and
propagation of forest fire \cite{clar}.

For a system exhibiting dynamical scaling, the r.m.s. roughness $W$ of
an interface is characterized by the following scaling with
respect to time
$t$ and the lateral system width $L$:
\[
W(L,t) \sim L^\alpha f(t/L^z),
\]
where the scaling function $f(u)$ behaves as $f(u) \sim u^\beta$ with
$\beta= z/\alpha$ for $u \ll 1$ and $f(u) \sim$ constant for $u \gg
1$. The exponent $\alpha$ is the roughness exponent that describes the
dependence of the saturated surface roughness versus the lateral
system size, while the exponent $\beta$ describes scaling at an early
stage when finite-size effects are negligible. The crossover time
between the two regimes is $t_s=L^{z}$.

A widely studied phenomenological equation representing the
non-equilibrium growth of such interfaces is the
Kardar-Parisi-Zhang (KPZ) equation. In $1+1$ dimensions, it states that:
\begin{equation}\label{eq.kpz}
    \frac{\partial h(x,t)}{\partial t}= \nu_0 \frac{\partial^2 h(x,t)}{\partial^2 x} + \lambda_0 \left(\frac{\partial h(x,t)}{\partial x}\right)^2+ \eta(x,t)
\end{equation}
where $h(x,t)$ is the local surface height at lateral coordinate $x$
and time $t$. The coefficients $\nu$ and $\lambda$ represent the strength
of the linear and nonlinear surface smoothing terms respectively.
The noise $\eta(x,t)$ is Gaussian with zero variance and
covariance
\[
\langle \eta(x,t)\;\eta(x^{'},t{'})\rangle = 2 D_0 \delta(x -x{'})\;
\delta (t - t{'})
\]
where $D_0$ is the strength of the noise. The exponents characterizing
the KPZ equation in the highly nonlinear strong coupling are
$\alpha=1/2$ and $\beta=1/3$. In contrast, at $\lambda=0$ the linear
Edward Wilkinson (EW) equation is recovered leading to the weak
coupling exponents $\alpha =1/2$ and $\beta=1/4$. When both $\nu$ and
$\lambda$ are zero the growth reduces to simple random deposition (RD)
with $\beta=1/2$ but a lack of any saturation regime.

There has been a recent interest in the study of competitive growth
models (CGM) analyzing the interplay and competition between two
current growth processes for a single surface. These CGM are often
more realistic in describing growing in real materials \cite{shapir},
in which more than one microscopic growth mode usually exist. For
example, two distinct growth phases were observed in experiments on
interfacial roughening in Hele-Shaw flows \cite{hs1,hs2} as well
as in simulations on electrophoretic deposition of polymer chains
\cite{ep1,ep2}.  The resulting universalities from these competing
processes are not well understood \cite{albano,albanojpa,lidia,reis}.
Recently Horowitz et al.  \cite{albanojpa} introduced a CGM called
BD/RD in which the microscopic growth rule follows either that of the
ballistic deposition (BD) model with probability $p$ or simple random
deposition (RD) with probability $1-p$. This system exhibits a
transition at a characteristic time from RD to KPZ.  They found
numerically that the scaling behavior of $W$ in that model is giving
by
the empirical form
\begin{equation}
\label{eq.scaleW}
W \sim \frac{L^\alpha}{p^{\delta}} F\left(\frac{ t}{p^{-y} L^z}\right)
\end{equation}
Based on numerical estimates, exact values
\begin{equation}\label{eq.dy1}
\delta=1/2   \mbox{ ~~~~ and ~~~~} y=1
\end{equation}
have been conjectured for the BD/RD model.  Based on this conjecture,
the authors concluded using scaling arguments that the model follows
Eq. (\ref{eq.kpz}) with $\nu \sim p$ and $\lambda \sim p^{3/2}$.
Subsequently, a similar CGM namely EW/RD describing the competition
between the EW model and RD \cite{albano}. The simulations showed that
Eq.  (\ref{eq.scaleW}) also holds with a different set of exponents
which are conjectured to be
\begin{equation}\label{eq.dy2}
\delta=1   \mbox{ ~~~~ and ~~~~} y=2
\end{equation}
This model can also be described by Eq. (\ref{eq.kpz}) with $\nu
\sim p^2$ and $\lambda = 0$ \cite{albano,lidia}.

In this paper, we explain the scaling form (\ref{eq.scaleW}) and
derive rigorously the exact exponents $\delta$ and $y$ using simple
arguments. In addition, from the above examples of CGM, one might be
tempted to conclude that a CGM based on RD and a model in the KPZ (EW)
universality class should always lead to an overall process in the KPZ
(EW) class. We suggest that these naive predictions of the
universality is not always correct. Close examination of the
microscopic details of the growth models is indeed essentially. This
is illustrated by introducing a constrained EW (CEW) model. Although
this model essentially belongs to the EW class, a CGM in the form
CEW/RD at sufficiently small $p$ results in an overall process in the
KPZ universality class. 
It also demonstrates that a CGM can crossover from one universality class to
another by varying  $p$.

\section{Exact scalings for CGMs}\label{S.exs}

In the RD model a particle is dropped at a randomly selected column
increasing the local surface height by one.  For the CGMs BD/RD or
EW/RD described above, a particle is deposited on the surface following a RD
process with probability $1-p$ and by another process $A$ (which is
either BD or EW) with probability $p$. Now, we derived analytically
the exact exponents $\delta$ and $y$ which characterize the $p$
dependence of the scaling behavior of $W$ given in
Eq.~(\ref{eq.scaleW}). We consider $p \rightarrow 0$. At each unit
time, $L$ particles are deposited. The average time interval between
any two consecutive $A$ events at any column $i$ is $\tau = 1/p$.
During this period, $\tau-1 \simeq \tau$ atoms on average are directly
stacked onto the surface according to the simple RD rule. The local
height at column $i$ hence increases by $\eta_i$ which is an
independent Gaussian variable with mean $\overline{\eta}=\tau$ and standard
deviation $\sigma_{\eta}= \sqrt{\tau}$ according to the central limit
theorem.  The mean however only leads to an irrelevant rigid shift of
the whole surface. We can easily apply a vertical translation so that
$\overline{\eta}=0$. 
%
After these $\tau$ RD events at column $i$, one $A$ event on average
is expected at the same column. 

Now we consider $A$ to be the BD process. The CGM is then
the BD/RD model. When a BD event occurs at column $i$, its height is
updated in the simulations according to $h_i \rightarrow max\{ h_{i-1}, h_{i+1}, h_i+1 \}$.
In the limit $p\rightarrow 0$ so that $\sigma_{\eta}
>> 1$, the height of the atom is negligible
compared with the increments due to the RD events.
The growth rule hence reduces to 
\begin{equation}
\label{eq.BDlimit}
h_i \rightarrow max\{ h_{i-1}, h_{i+1}, h_i \}
\end{equation} 
We have now arrived at a limiting BD/RD model defined as follow: At
every coarsened time step $\tau=1/p$, the local height $h_i$ at every
column $i$ first changes by an additive Gaussian noise term $\eta_i$
with mean zero and standard deviation $\sigma_\eta=\sqrt{\tau}$. Then
the limiting BD growth rule in Eq.  (\ref{eq.BDlimit}) is applied to
every column $i$. A more careful analysis should account for the fact
that the BD events at various columns indeed occur randomly and
asynchronously during the period $\tau$ but this will not affect our
result.  In this limiting model, the time and the vertical length
scales are determined completely by $\tau$ and $\sigma_\eta$
respectively. Therefore, time scales as $t \sim\tau = 1/p$ while
roughness scales as $W \sim \sigma_\eta \sim 1/p^{1/2}$.  This
explains the $p$ dependence of the scaling form in Eq.
(\ref{eq.scaleW}). In particular, we obtain the exact exponents $y=1$
and $\delta=1/2$ in agreement with values in Eq. (\ref{eq.dy1}) first
conjectured in Ref.  \cite{albano} but {\it not} derived analytically
before.

Next, we assume that $A$ represents the EW growth rule instead. For
this growth rule, a particle is dropped at a random column but when it
reaches the surface it is allowed to relax to the lower of the nearest
neighboring columns. If the heights at both nearest neighbors are
lower than the selected one the relaxation is directed to either of
them with equal probability. Our CGM now becomes a EW/RD model. The
corresponding derivation of the characteristic length and time scales
is similar to that for the BD/RD case. Assuming again $p \rightarrow
0$, the average time interval between any two consecutive EW events at
any given site is $\tau = 1/p$. Consider a characteristic time
$n\tau$. On average $n\tau$ RD events occur at any given site leading
height increments with a standard deviation $\sqrt{n\tau}$.  However,
only $n$ EW events take place. The resulting smoothing dynamics is
such that a big step for instance will typically decrease in height by
$n$. For scaling to hold, the two length scales $\sqrt{n\tau}$ and $n$
have to be identical and we obtain $\tau \sim n$. The characteristic
time scale considered is hence $n \tau \sim 1/p^2$, while a
characteristic length scale for the surface height is $\sqrt{n\tau}
\sim 1/p$. We have hence derived $y=2$ and $\delta=1$ previously
conjectured in Ref. ~\cite{albano}.

\section{CEW and CEW/RD models}
We now introduce the constrained EW model (CEW) which is a
generalization of the EW model. In $1+1$ dimensions, particles are
aggregated by the following rules.  We choose a site $i$ at random
among the $L$ possible sites. The surface height $h_i$ at the selected
column is increased by one if this height is lower than the values
$h_{i\pm 1}$ at the neighboring columns. Otherwise either $h_{i-1}$ or
$h_{i+1}$, whichever smaller, is increased to
\begin{equation}
\label{eq.cew}
h_{i \pm 1}=\max\{h_{i \pm 1}+1, h_i-c \}
\end{equation}
If $h_{i-1}=h_{i+1}$,
either one will be updated with equal probability.
Growth at ${i\pm 1}$ physically represents the rollover of a newly
dropped particle to a lower site nearby under the influence of gravity
for example. In the original EW model, there is no limit in the
vertical distance transversed during the rollover and the particle can
in principle slides down a very deep cliff if one exists next to
column $i$. This is unphysical if there is a finite chance for the
sideway sticking of the particle to the cliff. In the CEW growth rule
defined in Eq. (\ref{eq.cew}) , this vertical drop during rollover is
limited to $c$ by the process of sideway sticking.
As $c \rightarrow \infty$, it is easy to see that CEW reduces to the
standard EW model. At $c=0$, sideway sticking of particles occurs
frequently and CEW behaves similarly to BD, although the precise
growth rules are different. In the rest of this paper, we put $c=4$.
\begin{figure}[h]
\begin{center}
\includegraphics[width=5cm,height=5.8cm,angle=-90]{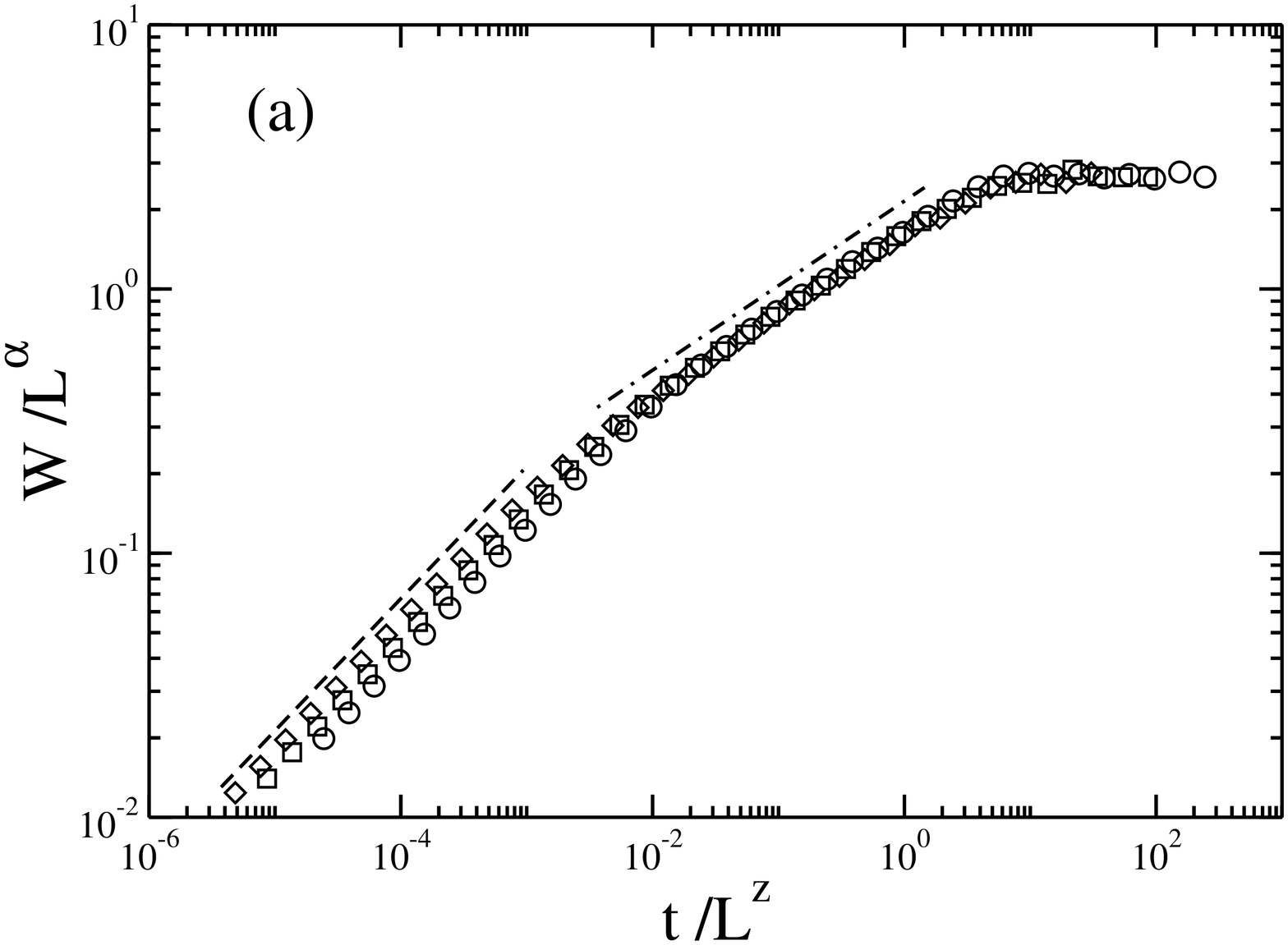}
\includegraphics[width=5cm,height=5.8cm,angle=-90]{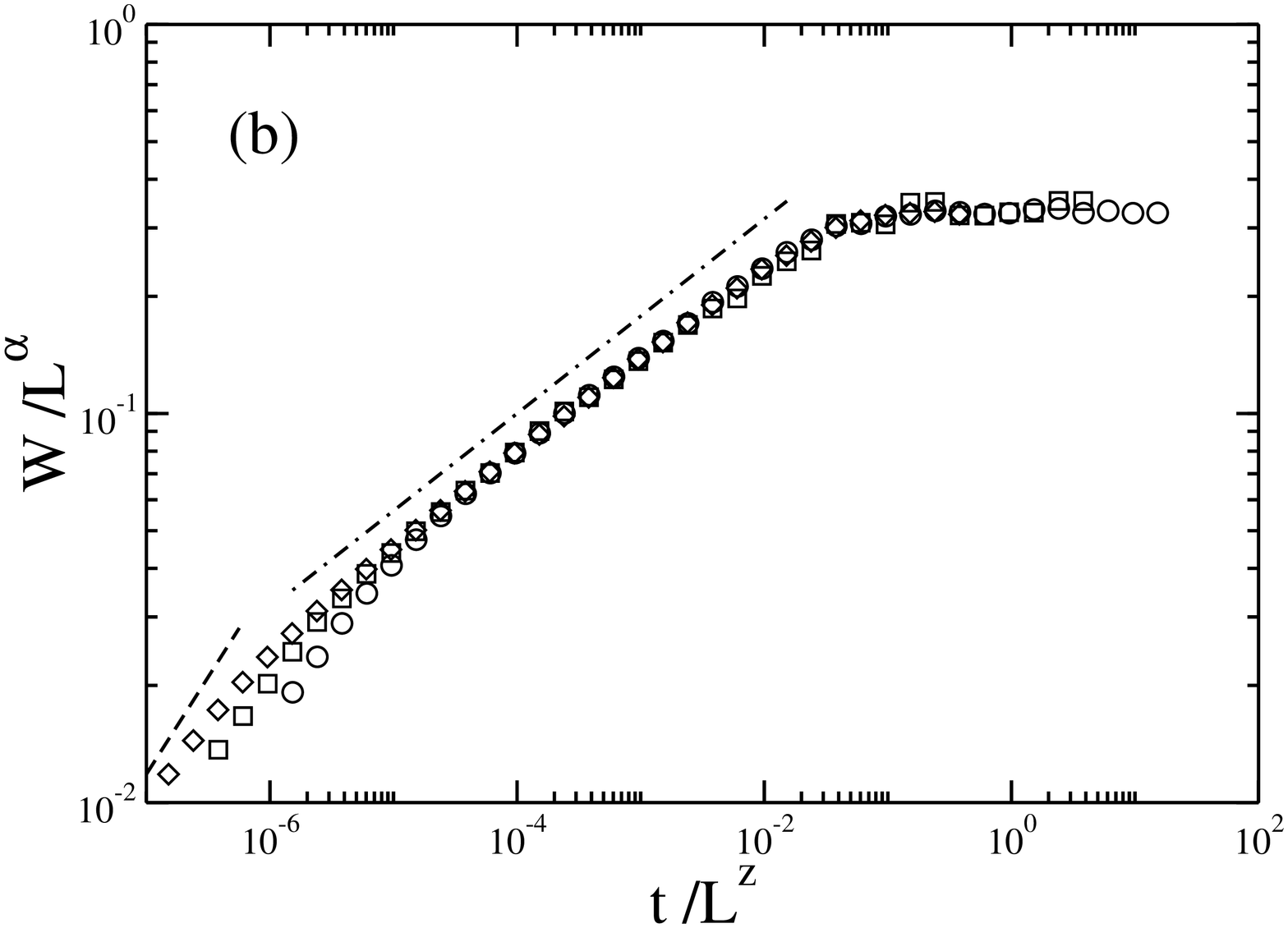}
\caption{Scaling plot of $ W /L^\alpha$ as function of $t/L^z$ for $c=4$ and different $L$ values:
$L= 256$ ($\circ$),
$L=512 $ ($\Box$),
$L=1024$ ($\diamond$).
(a) For $p=0.02$ we use the scaling exponents characteristic of the KPZ equation
$\alpha=1/2$ and $z=3/2$. The dashed line with slope $1/2$ and the dotted-dashed line with slope $1/3$ are used has a guide to show the RD 
and KPZ regimes respectively (b) Show the same as  (a) but for $p=0.64$, with $\alpha=1/2$ and $z=2$, characteristic of the EW.
behavior.  The dashed line with slope $1/2$ and the dotted-dashed line with slope $1/4$ are used has a guide to show 
the RD and the EW regimes respectively.
\label{f.1}}
\end{center}
\end{figure}
As will be demonstrated by the simulations presented
later, the CEW model at $c=4$ at practical length and time scales
belongs to the EW universality class.
Finally, we can define the CEW/RD model which is a CGM based on RD and
CEW. Similar to the definitions of other CGM
defined above, at each simulation step, the CEW growth rule is applied
with probability $p$ while a RD event occurs with probability $1-p$.
Time $t$ is then increased by $1/L$.

Now we present the simulation results. In Fig.~\ref{f.1}(a) and
Fig.~\ref{f.1}(b) we show the Log-Log plot of $W/L^\alpha$ as function
of $t /L^z$ for two limiting values of $p$. For $p \to 0$ the behavior
is consistent with the KPZ universality class with $\alpha=1/2$ and
$z=3/2$, while for $p \to 1$ the system behaves as predicted by the EW
equation with $\alpha=1/2$ and $z=2$. The initial regime corresponding
to the RD deposition, with $\beta=1/2$ does not scale with the system
size.  In order to show that the universality class depends
on $p$, we compute $\beta$ as function of $t$ using successive slopes
defined in Ref~\cite{lidiasaw}.
\begin{figure}[h]
\begin{center}
\includegraphics[width=7cm,height=8cm,angle=-90]{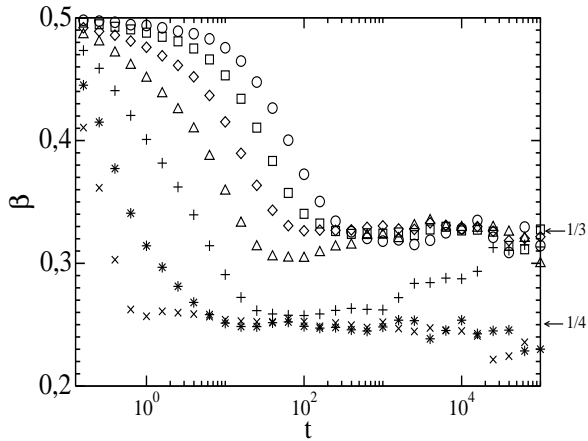}
\caption{Plot of the dynamic exponent  $\beta$ as function of $\log\; t$ for $L=8192$ and different $p$ values:
$p=0.02$ ($\circ$),
$p=0.04$ ($\Box$),
$p=0.08$ ($\diamond$),
$p=0.16$ ($\bigtriangleup$),
$p=0.32$ ($+$),
$p=0.64$ ($\ast$), and
$p=1.0$ (x) 
showing the change in the behavior of $\beta$ with time. The $\beta$ values where computed over 100 realizations.
The arrows are used as guides to show the asymptotic exponents\label{f.2}}
\end{center}
\end{figure}

Figure~\ref{f.2} plots $\beta$ as a function of $\log \;t$ for
different values of $p$. At the beginning $\beta= 1/2$ as expected for
the initial RD regime. After this early regime, the system evolves
either to the KPZ class with $\beta=1/3$ for $p \to 0$ or to the EW
class for $p \to 1$ with $\beta=1/4$. For intermediate $p$ values
(after the RD regime), the system behaves as in the weak coupling of
Eq.~(\ref{eq.kpz}). It is easy to observe a transition from an EW to a
KPZ for $p \ge 0.32$ while for $p \to 0$ the system always belongs to
the KPZ universality class.

\section{Generalized Continuous Equations and Scaling}\label{S.gce}
As we show above the CEW/RD model has a transition from a KPZ to a EW
as the tuning parameter $p$ goes from $0$ to $1$.  Thus the nonlinear
coefficient $\lambda$ of the KPZ equation has to vanish as $p \to 1$.
In order to understand the functional form of $\lambda(p)$ we perform
a finite size scaling analysis of the growth velocity based on 
\cite{meakin}:
\begin{equation}\label{eq.dv}
\Delta v(L) \sim \lambda L ^{-\alpha} ~~~~\hbox{for $t \gg t_s$}
\end{equation}  
where $\Delta v (L,t) = v(L=1024,t) -v(L=10,t)$ and $v(L,t) = \langle dh /dt \rangle$. The $\Delta v$ correction should go 
to zero when the nonlinear term $\lambda$ vanishes.
\begin{figure}[h]
\begin{center}
\includegraphics[width=7cm,height=7cm,angle=-90]{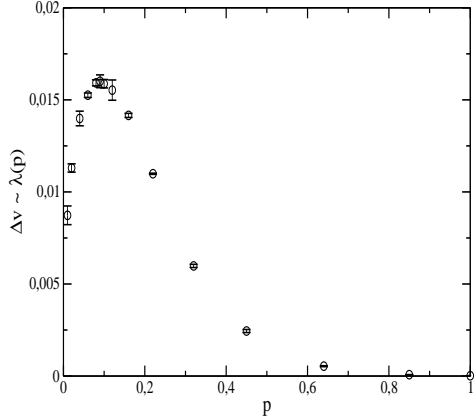}
\caption{Linear plot of $\lambda$ as function of $p$. The averages where taken over typically $500$ realizations.
The errors bars where computed using $5$ independent sets of $500$ realizations each. 
\label{f.lambda}}
\end{center}
\end{figure}
Thus, using Eq. (\ref{eq.dv}) we can determine how $\lambda$ in the
KPZ equation depends on $p$.  In Fig~.\ref{f.lambda} we plot $\Delta
v$ as function of $p$ for fixed $L$ to show the $p$ dependence of
$\lambda$. From the plot we can see that the functional form for the
CEW/RD CGM is totally different from the scaling form $\lambda(p) \sim
p^{3/2}$ of the BD/RD model \cite{albanojpa}.  For small $p$ values
$\lambda$ has a power law dependence with $p$, while for $p \to 1$,
$\lambda(p) \to 0$ with a fast decay. 
Now we proceed to
generalize Eq.~(\ref{eq.kpz}) for the CGM model applying the following
transformation :  $h^{'}= h \; f(p) \;b^\alpha$, $x^{'}= b \;x $ and 
$t^{'} = g(p)\; t \;b ^z$.
where $b$ is the length transformation. As the interface evolution of these
CGM are independent of $b$ they can be described, after applying the transformations defined above, by the generalized continuous equation:
\begin{equation}\label{eq.dh}
    \frac{d h(x,t)}{dt}= \nu(p) \frac{\partial^2 h(x,t)}{\partial^2 x} + \lambda(p) \left(\frac{\partial h(x,t)}{\partial x}\right)^2+ D(p)\;\eta(x,t)
\end{equation}
with
\begin{eqnarray}\label{eq.coefp1}
  \nu(p) & = & \nu_0 \; g(p) \\\label{eq.coefp2}
  \lambda(p) & =& \lambda_0\; f(p) \;g(p)\\
  D(p) & =& \frac{g(p)}{f(p)^2} \label{eq.coefp3}
\end{eqnarray}
where $g(p)$ is related to the characteristic time scale when the
correlations begins to dominate the dynamic of the interface and $f(p)$ is related
to the saturation length scale. Notice that for the EW universality class $\lambda_0 =0$
Using the exact results from Sect~\ref{S.exs}, 
\begin{eqnarray}\label{eq.f}
f(p) \sim \left\{%
\begin{array}{cc}
p^{1/2}, & \hbox{for BD/RD and CEW/RD when $p \to 0$, }\\ 
p , & \hbox{for CEW/RD when $p \to 1$, }\\
\end{array}
\right .
\end{eqnarray}
and
\begin{eqnarray} \label{eq.g}
g(p) \sim \left\{%
\begin{array}{cc}
p, & \hbox{for BD/RD and CEW/RD when $p \to 0$,}\\
p^2 , & \hbox{for CEW/RD when $p \to 1$, }\\
\end{array}
\right .
\end{eqnarray}
Thus Eqs.~(\ref{eq.coefp1}),(\ref{eq.coefp2}) and (\ref{eq.coefp3})
can be replaced by:
\begin{eqnarray}
   \nu(p) \sim \left\{%
\begin{array}{cc}
\nu_0 \; p, & \hbox{for BD/RD and CEW/RD when $p \to 0$, }\\
\nu_0 \;p^2 , & \hbox{for CEW/RD when $p \to 1$,}\\
\end{array}
\right .
\end{eqnarray}
\begin{eqnarray}
 \lambda(p) \sim \left\{%
\begin{array}{cc}
\lambda_0 \; p^{3/2}, & \hbox{for BD/RD when $p \to 0$,}\\
 0 , & \hbox{for CEW/RD  when $p \to 1$,}\\
\end{array}
\right .
\end{eqnarray}
and  $D(p) \sim  D_0$ for BD/RD and CEW/RD independant of $p$
\begin{figure}
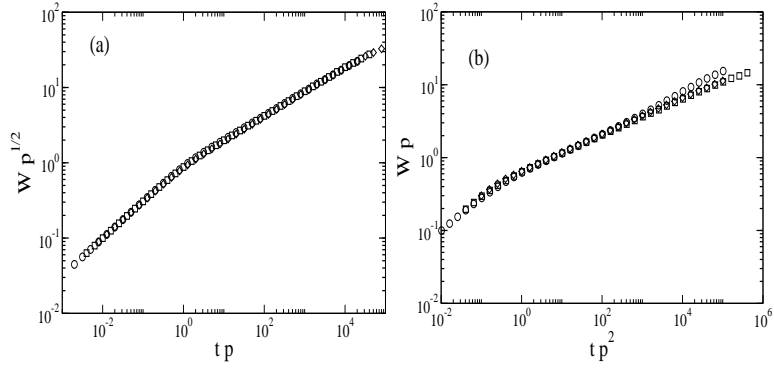

\begin{center}
\includegraphics[width=5cm,height=4.8cm]{Fig3a.eps}
\includegraphics[width=5cm,height=4.8cm]{Fig3b.eps}
\caption{Log- Log plot of $W \;f(p)$ as function of $t \;g(p)$ for $L=8192$ and different $p$ values.
In (a) $p=0.02$ ($\circ$), $p=0.04$ ($\Box$) and $p=0.08$ ($\diamond$), $g(p)=p$ and $f(p)=p^{1/2}$.
In (b) $p=0.32$ ($\circ$), $p=0.64$ ($\Box$) and $p=1.0$ ($\diamond$),  $g(p)=p^2$ and $f(p)=p$. 
Notice the departure from the EW scaling behavior for $p=0.32$
\label{f.3}}
\end{center}
\end{figure}
Thus, for the CGM belonging to the KPZ (EW) universality class the evolution of the interface 
is given by:
\begin{eqnarray} \label{eq.coef}
\frac{dh}{dt} = \left\{%
\begin{array}{cc}
\nu_0 \;p \;\frac{\partial^2 h(x,t)}{\partial^2 x} + \lambda_0\; p^{3/2} \left(\frac{\partial h(x,t)}{\partial x}\right)^2+ \eta_i(t), & \hbox{for BD/RD and CEW/RD when $p \to 0$,}\\
\nu_0 \;p^2\;\frac{\partial^2 h(x,t)}{\partial^2 x} +\eta_i(t) , & \hbox{for CEW/RD when $p \to 1$, }\\
\end{array}
\right .
\end{eqnarray}
The scaling behavior of $W$\cite{lidia}  is then:
\begin{eqnarray} \label{eq.scaling}
W f(p) / L^{\alpha} \sim\left\{%
\begin{array}{cc}
F\left(g(p) \lambda_0 \sqrt{\frac{D_0}{\nu_0}} \frac{t}{L^z} \right), & \hbox{for KPZ,}\\
F \left( g(p)\; \nu_0 \;\frac{t}{L^z}\right), & \hbox{for EW,}\\
\end{array}
\right .
\end{eqnarray}
that after replacing $f(p)$ and $g(p)$ given by Eq.~(\ref{eq.f}) and
Eq.~(\ref{eq.g}) leads to the exact scaling of $W$ predicted by
Eq.~(\ref{eq.scaleW}) with the exact values of $\delta$ and $y$
derived in Sec~\ref{S.exs}.  In Fig.~\ref{f.3} we show the Log-Log
plot of $W f(p)$ as function of $t \;g(p)$ in the two limiting $p$
values for fixed $L$.
The results are in agreement with our exact results (See
Sec~\ref{S.exs}) and our scaling ansatz (Eq~.(\ref{eq.scaling})).  

\section{Conclusions}
We derive analitically the $p$ dependance in the scaling behavior in
two CGMs named BD/RD and EW/RD.  Exact scaling exponents are derived
and are in agreement with previously conjectured values
\cite{albano,albanojpa}.  To our knowledge these exact scaling
behaviors were not analytically derived before.  This derivation
allows us to compute the scaling behaviors of the coefficients of the
continous equations that describe their universality classes.  We
introduce the CEW/RD model to show that not all CGMs can be
represented by an unique universality class.  The CEW/RD is an EW in
the limit $p \to 1$ while in the other limit $p \to 0$ it followings
the strong coupling behavior the KPZ equation. Our simulation results are in excellent agreement with the analytic
predictions.

\hspace{2cm}

{\bf Acknowledgments}: L.A.B.  thanks UNMdP for the financial support.
C.H.L. was supported by HK RGC, Grant No. PolyU-5289/02P.

\end{document}